\documentclass[10pt,journal,compsoc]{IEEEtran}
\ifCLASSOPTIONcompsoc
  \usepackage[nocompress]{cite}
\else
  \usepackage{cite}
\fi
\ifCLASSINFOpdf
   \usepackage[pdftex]{graphicx}
\else
\fi
\usepackage{booktabs}
\usepackage{amsmath}

\begin{document}
%
\title{An empirical approach to the relationship between emotion and music production quality}
%
%
%
%

\author{David~Ronan,
        Joshua D.~Reiss
        and~Hatice~Gunes
\IEEEcompsocitemizethanks{\IEEEcompsocthanksitem D. Ronan is with the Centre for Intelligent Sensing, Queen Mary University of London, UK.\protect\\
E-mail: d.m.ronan@qmul.ac.uk
\IEEEcompsocthanksitem J.D. Reiss is with the Centre for Digital Music, Queen Mary University of London, UK.\protect\\
E-mail: joshua.reiss@qmul.ac.uk
\IEEEcompsocthanksitem H. Gunes is with the Computer Laboratory, University of Cambridge, UK.\protect\\
E-mail: hatice.gunes@cl.cam.ac.uk }
}

%
%

\markboth{Journal of \LaTeX\ Class Files}%
{Shell \MakeLowercase{\textit{et al.}}: An empirical approach to the relationship between emotion and music production quality}
%



\IEEEtitleabstractindextext{%
\begin{abstract}
In music production, the role of the mix engineer is to take recorded music and convey the expressed emotions as professionally sounding as possible. We investigated the relationship between music production quality and musically induced and perceived emotions. A listening test was performed where 10 critical listeners and 10 non-critical listeners evaluated 10 songs. There were two mixes of each song, the low quality mix and the high quality mix. Each participant’s subjective experience was measured directly through questionnaire and indirectly by examining peripheral physiological changes, change in facial expressions and the number of head nods and shakes they made as they listened to each mix. We showed that music production quality had more of an emotional impact on critical listeners. Also, critical listeners had significantly different emotional responses to non-critical listeners for the high quality mixes and to a lesser extent the low quality mixes. The findings suggest that having a high level of skill in mix engineering only seems to matter in an emotional context to a subset of music listeners.
\end{abstract}

}

\maketitle

\IEEEdisplaynontitleabstractindextext

%
\IEEEpeerreviewmaketitle

\IEEEraisesectionheading{\section{Introduction}\label{sec:introduction}}
\IEEEPARstart{T}{here} are a number of stages when it comes to producing music for mass consumption. The first step is to record a musical performance using specific microphone placement techniques in a suitable acoustic space. In the post-production stage, the mix engineer combines the recordings through mixing and editing to achieve a final mix. Predominately, the more skilled the mix engineer is, the better the final mix sounds in terms of production quality. The mixing of audio involves applying signal processing techniques to each recorded audio track whereby the engineer manipulates the dynamics (balance and dynamic range compression), spatial (stereo or surround panning and reverberation), and spectral (equalisation) characteristics of the source material. Once the final mix has been created, it is sent to a mastering studio where  additional processing is applied before it can be distributed for listening in a home or a club environment \cite{case2011mix}.

There have been several studies that have looked at why people prefer certain mixes over others.  \cite{deman2015perceptual, deman2014d} conducted a mix experiment where groups of nine mix engineers were asked to mix 10 different songs. The mixes were evaluated in a listening test to infer the quality as perceived by a group of trained listeners. Mix preference ratings were correlated with a large number of low level features in order to explore if there was any relationship, but the findings indicated in this particular case was that there was no significantly strong correlations. 

In \cite{ronan2015automatic}, we analysed the same tracks used in \cite{deman2015perceptual, deman2014d}, to ascertain the impact of subgrouping practices on mix preference, where subgrouping involves combining similar instrument tracks for processing and manipulation. We looked at the quantity of subgroups and the type of subgroup effect processing used for each mix, then correlated these findings with mix quality ratings to see the extent of the relationship \cite{ronan2015impact}. 

\cite{wilson2015perception} claimed that audio production quality is linked to perceived loudness and dynamic range compression. It also demonstrated that a participant's expertise is not a strong factor in assessing audio quality or musical preference.

To our knowledge, there have been no previous studies that examined the relationship between music production quality and emotional response. This represents a new area of research in music perception and emotion that we intend to explore. In \cite{pestana2014intelligent}, three of the mix engineers that were interviewed mentioned the importance of emotion in the context of mixing and producing music. This indicates that emotion plays a significant role in how a mix engineer tries to achieve a desired mix. \cite{ross2007rest} states that dynamic contrast in a piece of music has been heralded as one of the most important factors for conveying emotion.

The purpose of the current study is to determine the extent of the link between music production quality and musically induced and perceived emotions. The participants in this study listened to low and high quality mixes (rated in \cite{deman2015perceptual, deman2014d}) of the same musical piece. We then measured each participant's subjective experience, peripheral physiological changes, changes in facial expressions and head nods, and shakes as they listened to each mix.

The rest of the paper is organised as follows. Section~\ref{sec:background} provides the background to this study with respect to musically induced vs. perceived emotions, psychological emotion models and measuring emotional responses to music. Section~\ref{sec:methodology} provides the methodology used to conduct this experiment. Section~\ref{sec:analysisandresults} presents the results and subsequent analysis. Section~\ref{sec:discussion} discusses the results, Section~\ref{sec:futurework} proposes future work and the paper is concluded in Section~\ref{sec:conclusion}.

\section{Background} \label{sec:background}

\subsection{Musically Induced vs. Perceived Emotions}
In the study of emotion and music listening, induced emotions are those experienced by the listener and perceived emotions are those conveyed in the music, though perceived emotions may also be induced \cite{gabrielsson2002emotion, eerola2010comparison, song2016perceived}. A listener's perception of emotional expression is mainly related to how they perceive and think about a musical process, in contrast to their emotional response to the music where someone experiences an emotion \cite{song2016perceived}.

Perceived emotion in music can be provoked in a number of ways. It can be associated with the metrical structure of the music, or how a certain song might be perceived as happy or sad because of the chords being played \cite{gabrielsson2002emotion}. Numerous studies have shown that any increase in tempo/speed, intensity/loudness or spectral centroid causes higher arousal. These studies have been summarised in \cite{gabrielsson2010role}, where tempo, loudness and timbre were shown to have an impact on how other typical `musical' variables such as pitch and the major-happy minor-sad chord associations are perceived.    

The most complete framework of psychological mechanisms for emotional induction is in \cite{juslin2008emotional} and its extensions \cite{juslin2010does, juslin2013everyday}.  Until that point, most research in that area had been exploratory, but Juslin et al. posited a theoretical framework of eight different cognitive mechanisms known as BRECVEMA. The eight mechanisms are as follows:

\begin{itemize}
 \item \textbf{Brain stem reflex} is a hard-wired primordial response that humans have to sudden loud noises and dissonant sounds. A reason given for the brain stem reflex reaction is the dynamic changes in music \cite{juslin2013everyday}. This particular mechanism might be related to music production in terms of a recording having good dynamics. A mix that has sudden large bursts in volume should arouse the listener more.
  
  \item \textbf{Rhythmic entrainment} is when the listener's internal body rhythm adjusts to an external source, such as a drum beat. This may relate to music production in a similar way as the brain stem reflex, i.e., if the drums in a musical production are loud and have a clear pulse, the listener may be more aroused.
  
  \item \textbf{Evaluative conditioning} occurs because a piece of music has been paired repeatedly with a positive or negative experience and an emotion is induced. 
  
  \item \textbf{Emotional contagion} is when the listener perceives an emotional expression in the music and mimics the emotions internally \cite{juslin2013makes}. This may mean that a better quality mix conveys the emotion in music in a clearer sense than a poorer quality mix, e.g. vocals or lead guitar is more audible in one mix over the other.
  
  \item \textbf{Visual imagery} may occur when a piece of music conjures up a particularly strong image. This could potentially have negative or positive valence and has been linked to feelings of pleasure and deep relaxation \cite{juslin2013everyday}.

  \item \textbf{Episodic memory} is when music triggers a particular memory from a listener's past life. When a memory is triggered, so is an attached emotion \cite{juslin2008emotional}. A mix engineer might use a certain music production technique from a specific era, which may trigger nostalgia in the listener.
  
  \item \textbf{Musical expectancy} is believed to be activated by an unexpected melodic or harmonic sequence. The listener will expect musical structure to be resolved, but suddenly it is violated or changes in an unexpected way \cite{juslin2013makes}. 
  
  \item \textbf{Aesthetic judgement} is the mechanism that induces `aesthetic emotion' such as admiration and awe. This may play a part in music production quality by enhancing musically induced emotions. How well a song has been mixed can be judged on the artistic skill involved as well as how much expression is in the mix. A poor mix is not typically going to be as expressive as a well constructed mix.   
\end{itemize}

How both perceived and induced emotions in music relate to music production quality is an area of music and emotion that has not yet been explored. For both induced and perceived musical emotions we have proposed a number of ways in which a mix engineer may have a direct effect, which we seek to capture from the listener through self-report, physiological measures, facial expression and body movement.

\subsection{Psychological Models of Emotion }
To describe musical emotions, three well-known models may be employed; discrete, dimensional and music specific. 

The discrete or categorical model is constructed from a limited number of universal emotions such as happiness, sadness and fear \cite{ekman1992argument, panksepp1998affective}. One criticism is that the basic emotions in the model are unable to describe many of the emotions found in everyday life and there is not a consistent set of basic emotions \cite{eerola2009prediction,sloboda2001psychological}.

Dimensional models consider all affective terms along broad dimensions. The dimensions are usually related to valence and arousal, but can include other dimensions such as pleasure or dominance \cite{russell1980circumplex, mehrabian1995framework}. Dimensional models have been criticised for blurring the distinction between certain emotions such as anger and fear, and because participants can not indicate they are experiencing both positive and negative emotions \cite{eerola2009prediction, song2016perceived, sloboda2001psychological}.

In recent years, a music-specific multidimensional model has been constructed. This is derived from the Geneva Emotion Music Scale (GEMS) and has been developed for musically induced emotions. This consists of nine emotional scales; wonder, transcendence, tenderness, nostalgia, peacefulness, power, joyful activation, tension and sadness \cite{zentner2008emotions,song2016perceived}. The scales have been shown to factor down to three emotional scales; calmness-power, joyful activation-sadness and solemnity-nostalgia \cite{zentner2008emotions, pearce2015age}. 

Empirical evidence \cite{kreutz2007using, vieillard2008happy} suggests both discrete and dimensional models are suitable for measuring musically induced and perceived emotions \cite{song2016perceived}. \cite{zentner2008emotions} compared the discrete approach, the dimensional approach and the GEMS approach. It was found that participants preferred to report their emotions using the GEMS approach. Therefore, we adopted the GEMS approach as well as the dimensional model.

\subsection{Measuring Emotional Responses to Music}
We employed self-report, physiological measures, facial expression analysis and head nod-shake detection for measuring emotional responses to music. 

\subsubsection{Self-Report Methods}
The most common self-report method to measure emotional responses to music is to ask listeners to rate the extent to which they perceive or feel a particular emotion, such as happiness. Techniques to assess affect are measured using a Likert scale or choosing a visual representation of the emotion the person is feeling. An example visual representation is the Self-Assessment Manikin \cite{bradley1994measuring} where the user is asked to rate the scales of arousal, valence and dominance based on an illustrative picture.

Another method is to present listeners with a list of possible emotions and ask them to indicate which one (or ones) they hear. Examples are the Differential Emotion Scale and the Positive and Negative Affect Schedule (PANAS). In PANAS, participants are requested to rate 60 words that characterise their emotion or feeling. The Differential Emotion Scale contains 30 words, 3 for each of the 10 emotions. These would be examples of the categorical approach mentioned previously \cite{izard2007basic, watson1988development}. 

A third approach is to require participants to rate pieces on a number of dimensions. These are often arousal and valence, but can include a third dimension such as power, tension or dominance \cite{grewe2007emotions, eerola2009prediction}.

Self-reporting leads to concerns about response bias. Fortunately, people tend to be attuned to how they are feeling (i.e., to the subjective component of their emotional responses) \cite{hunter2010music}. Furthermore, Gabrielsson came to the conclusion that self-reports are ``the best and most natural method to study emotional responses to music'' after conducting a review of empirical studies of emotion perception \cite{gabrielsson2002emotion}. One caveat with retrospective self-report is `duration neglect' \cite{schubert2010continuous}, where the listener may forget the momentary point of intensity of the emotion attempted to be measured.

We chose self-report in our experiment due to it being the most reliable measure according to \cite{gabrielsson2002emotion}. GEMS-9 was used for measuring induced emotion and Arousal-Valence-Tension for perceived emotion. We selected GEMS-9 due to it being a specialised measure for self-report of musically induced emotions and Arousal-Valence-Tension due to it being a dimensional rather than categorical model.

\subsubsection{Physiological Measures}
Measures for recording physiological responses to music include heart or pulse rate, galvanic skin response, respiration or breathing rate and facial electromyography. Such measures have been used in recent papers \cite{egermann2013probabilistic, juslin2013makes, morgan2015using}. 

High arousal or stimulative music tends to cause an increase in heart rate, while calm music tends to cause a decrease \cite{hodges2010psychophysiological}. Respiration has been shown to increase in 19 studies on emotional responses to music \cite{hodges2010psychophysiological}. These studies found differences between high- and low-arousal emotions but few differences between emotions with positive or negative valence. 

One physiological measure that corresponds with valence is facial electromyography (EMG). EMG measurements of cheek and brow facial muscles are associated with processing positive and negative events, respectively \cite{schwartz1980facial}. In \cite{witvliet2007play}, each participant's facial muscle activity was measured while they listened to different pieces of music that were selected to cover all parts of the valence-arousal space. Results showed greater cheek muscle activity when participants listened to music that was considered high arousal and positive valence. Brow muscle activity increased in response to music that was considered to induce negative valence, irrespective of the arousal level. 

\textit{Galvanic skin response} (GSR)  is a measurement of electrodermal activity or resistance of the skin \cite{andreassi2013psychophysiology}. When a listener is aroused, resistance tends to decrease and skin conductance increases \cite{lundqvist2008emotional, roy2009modulation}. We used skin conductance measurements in our experiment as it has been used extensively in previous studies related to music and emotion \cite{hodges2010psychophysiological,egermann2013probabilistic,juslin2013makes,morgan2015using}. 

\subsubsection{Facial Expression and Head Movement}

The Facial Action Coding System (FACS) \cite{hager2002facial} provides a systematic and objective way to study facial expressions, representing them as a combination of individual facial muscle actions known as Action Units (AU). Action Units can track brow and cheek activity, which can be linked to arousal and valence when listening to music \cite{witvliet2007play}.

\cite{weisgerber2015facial} examined how schizophrenic patients perceive emotion in music using facial expression, and \cite{silvey2012role} looked at the role of a musical conductor’s facial expression in a musical ensemble. We were unable to find anything directly related to our research questions.

People move their bodies to the rhythms of music in a variety of different ways. This can occur through finger and foot tapping or other rhythmic movements such as head nods and shakes \cite{wallin2001origins, sedlmeier2011music}. In human psychology, head nods are typically associated with a positive response and head shakes  negative one \cite{tom1991role}. In one study, participants who gauged the content of a simulated radio broadcast more positively were more inclined to nod their head than those who performed a negatively associated head shaking movement \cite{wells1980effects, sedlmeier2011music}. But for music, a head shake might be considered a positive response as this might simply be a rhythmic response.

We examined facial expression in this experiment since it had not been attempted before in music and emotion or music production quality research. Facial expression analysis is somewhat similar to facial EMG, so we should be able to link results to previous findings \cite{hodges2010psychophysiological}.

\section{Methodology} \label{sec:methodology}

\subsection{Research questions and hypotheses}
Our original hypothesis was that music production quality had a direct effect on the induced and perceived emotions of the listener. However, before we proceeded to the main study, we conducted a short pilot study on six participants, three of whom had critical listening skills. The feedback from the pilot study indicated that training was required in order for participants to become familiar with the adjectives used to describe induced emotions. We also decided to track head nods and shakes, a typical response to musical enjoyment, based on a review of the recorded videos. Observation of potential differences between critical and non-critical listeners led us to revise our original hypothesis. It was refined to be that music production quality has more effect on the induced and perceived emotions of critical listeners than non-critical listeners.

\subsection{Participants}
Twenty participants were recruited from within the university. 14 were male, 6  female and their ages ranged from 26 to 42 ($\mu = 30.4, \sigma^2 = 4.4$). 10 participants had critical listening skills, i.e, knew what critical listening involved and had been trained to do so previously or had worked in a studio, while the other 10 did not i.e., no music production experience and not trained in how to critique a piece of music. A pre-experiment questionnaire established the genre preference of participants, shown in Table~\ref{genrepref}, since some participants may have bias towards certain genres.

\begin{table}[ht]
\centering
\caption{Genre preference for participants}
\label{genrepref}
\begin{tabular}{@{}cc@{}}
\toprule
\textbf{Genre}   & \textbf{No. of Participants} \\ \midrule
Rock/Indie       & 15                       \\
Dance/Electronic & 11                       \\
Pop              & 8                        \\
Jazz             & 6                        \\
Classical        & 4                        \\ \bottomrule
\end{tabular}
\end{table}
\vspace{-0.5cm}
\subsection{Stimuli}
Ten different songs were used, each with nine mixes (90 mixes in total). Songs were split into three study groups, where mixes for songs within a study group were created by 8 student mix engineers and their instructor, who was a professional mix engineer (the same professional mix engineer participated in Groups 1 and 2). These mixes were obtained from the experiment conducted in \cite{deman2015perceptual}. Mixes of a song had been rated for mix quality by all the members of the other study groups, so no one rated their own mix. Further details on how the stimuli was obtained can be seen in \cite{deman2015perceptual}. For our experiment, we selected the lowest and highest quality mix of each song. Table~\ref{tab:songsused} shows the names of each song, the song genre and which group mixed each song. Some song names had to be removed due to copyright issues, but the rest are available on the Open Multitrack Testbed \cite{deman2014omtb}. All mixes were loudness normalised using ITU-R BS. 1770-2 specification \cite{itu2011itu} to avoid bias towards loud mixes.

\begin{table}[ht]
\centering
\caption{Song titles, song genres and mix groups. Songs in italics
are not available online due to copyright restrictions.}\label{tab:songsused}
\begin{tabular}{@{}lll@{}}
\toprule
\textbf{Song Name} &   \textbf{Genre}   & \textbf{Mixed By} \\ \midrule
Red to Blue - (S1) & Pop-Rock   & Group 1    \\
Not Alone - (S2)  &  Funk      &Group 1     \\
\textit{My Funny Valentine} - (S3) & Jazz & Group 1    \\
Lead Me - (S4)  & Pop-Rock   & Group 1    \\
In the Meantime - (S5) &  Funk   & Group 1    \\
 - (S6) & Soul-Blues & Group 2     \\
\textit{No Prize} - (S7) &     Soul-Jazz       & Group 2    \\
 - (S8) &   Pop-Rock   & Group 2     \\
Under a Covered Sky - (S9) & Pop-Rock  &Group 2     \\
Pouring Room - (S10) &  Rock-Indie      & Group 3    \\ \bottomrule
\end{tabular}
\end{table}
\vspace{-0.5cm}
\subsection{Measurements}
\subsubsection{Physiological Measures}
To measure skin conductance we used small (53mm x 32 mm x 19 mm) wireless GSR sensors developed by Shimmer Research. The GSR module was placed around the wrist of their usually inactive hand, and electrodes strapped to their index and middle finger. ECG measurements were attempted but discarded due to extreme noise levels in the data, at least partly since participants moved in the rotatable chair provided. 

\subsubsection{Facial Expression and Head Nod-Shake}
\begin{figure}[ht]
\includegraphics[scale=0.295]{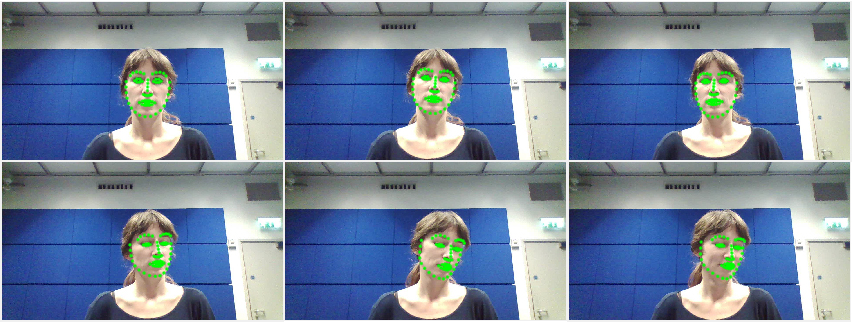}
\caption{\label{facialactionusits} Facial features tracked for detecting facial action units during music listening.}
\end{figure}

To record video for facial expression and head nod/shake detection, we used a Lenovo 720p webcam that was embedded in the laptop used to perform the experiment. In Figure~\ref{facialactionusits} we can see the automatic facial feature tracking for one of our participants.

%
%
\vspace{-0.5cm}
\subsubsection{Self-Report}
After listening to each piece of music, participants GEMS-9 to rate the emotions induced while listening. This was done using a 5-point Likert scales ranging from `Not at all' to `Very much' based on 9 adjectives; wonder, transcendence, power, tenderness, nostalgia, peacefulness, joyful activation, sadness and tension. Each participant also rated the emotions they perceived in each song using three discrete (1-100) sliders for arousal, valence and tension. They were also asked to indicate how much they liked each piece of music they heard based on a 5-point Likert scale ranging from `Not at all' to `Very much'.

\subsubsection{User Interface}
The physiological measurements, self-report scores and video were recorded into a bespoke software program developed for the experiment. It was designed to allow the experiment to run without the need for assistance, and the graphical user interface was designed to be as aesthetically neutral as possible. 


\subsubsection{Pre- and Post-Experiment Questionnaires}
We provided pre- and post-experiment questionnaires. The pre-experiment questionnaire asked simple questions related to age, musical experience, music production experience, music genre preference and critical listening skills. There was also a question clarifying each participant's emotional state as well as how tired they were when they started the study. If any participant indicated that they were very tired, we asked them to attempt the experiment at a later time once rested. 

The post-experiment questionnaire asked questions such as could they hear an audible difference between the two mixes of each song, was there any difference in emotional content between the two mixes of each song and was there any difference in the induced emotions between the two mixes of each song. These were all asked on a 5-point Likert scale ranging from `Not at all' to `Very much'.

\subsection{Setup}
\begin{figure}[ht]
\centering
\includegraphics[scale=0.055]{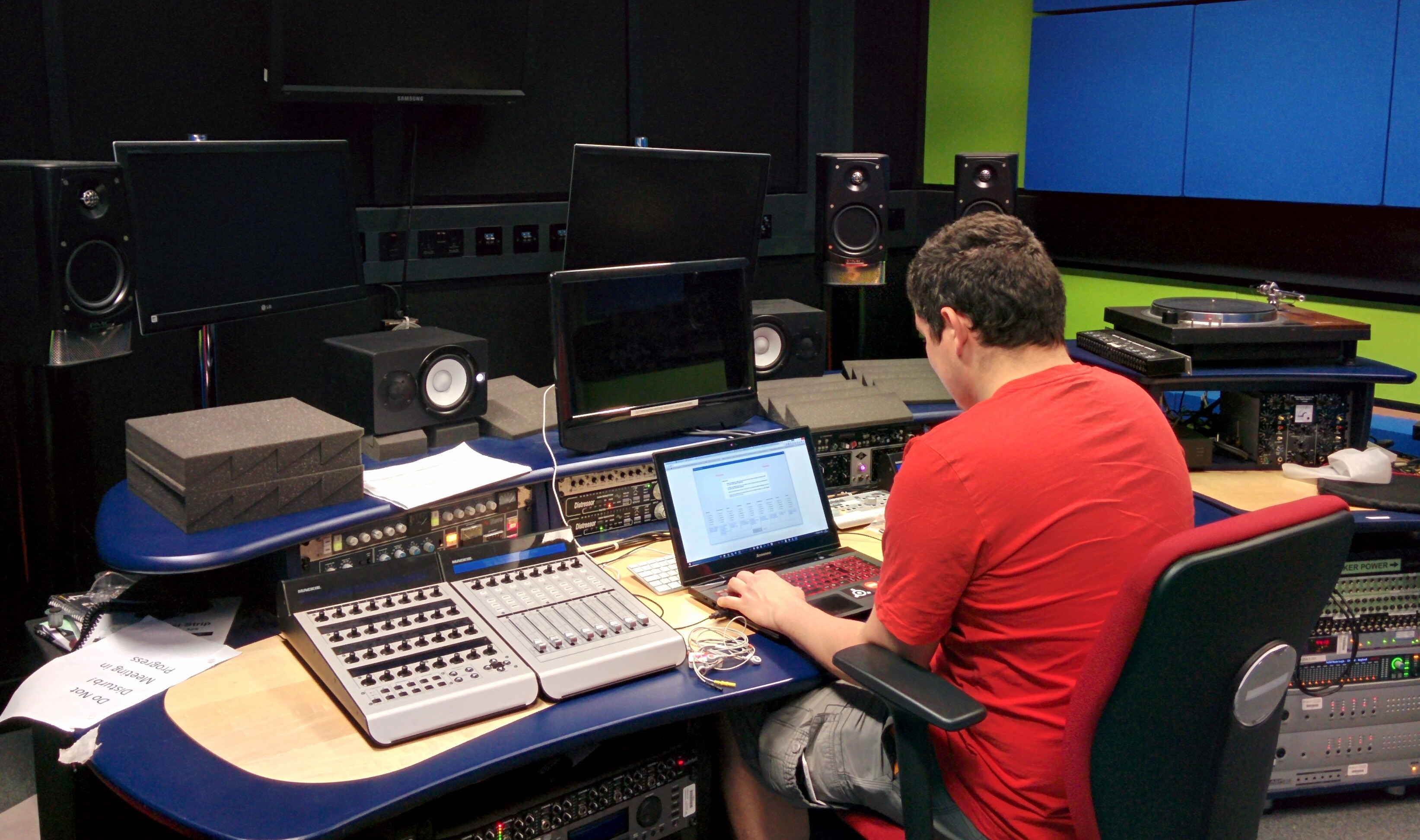}
\caption{\label{room} Studio space where the experiment was conducted.}
\end{figure}

The experiment took place in a dedicated listening room at the university. The room was very well lit, which was important for facial expression analysis and head nod/shake detection. Each participant was sat at a studio desk in front of the laptop used for the experiment. The audio was heard over a pair of studio quality loudspeakers, where the participant could adjust the volume of the audio to a comfortable level. Figure~\ref{room} shows the room in which the experiment was conducted.

\subsection{Tasks}
After the pre-experiment questionnaire, we trained each participant in how the interface worked. They were supervised while they listened to two example songs and were shown how to answer each question. 

\begin{figure*}[ht]
\centering
\includegraphics[scale=0.46]{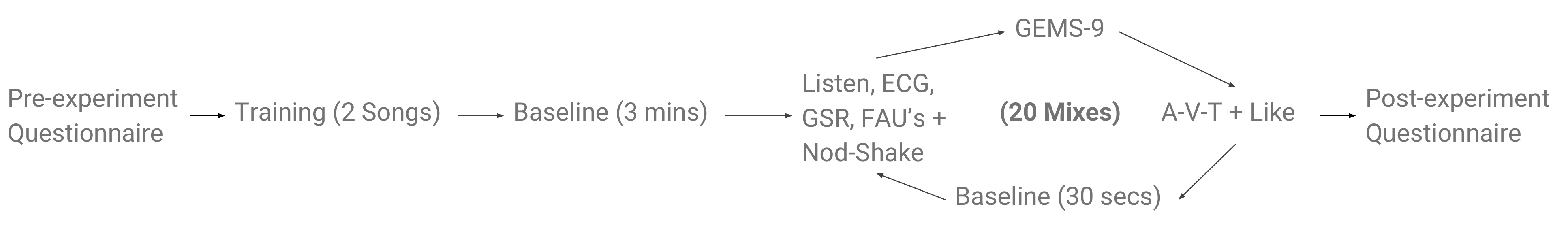}
\caption{\label{experimentprocess} Tasks involved in the experiment.}
\end{figure*}

Each participant was then asked to relax and listen to the music as they would at home for enjoyment. Next, three minutes of relaxing sounds were played to each participant in order to get an emotional baseline. They then had to click play in order for one of the mixes to be heard, where the order in which mixes were presented was randomised. While the music was playing,  GSR measurements and facial and head movements were recorded. Once the music finished, each participant rated the induced emotions using GEMS-9. They then rated perceived emotions on the Arousal-Valence-Tension scale and rated how much they liked each mix. Once answers were submitted, there was another 30 seconds of relaxing sounds played for an emotional baseline and the same procedure repeated for the next mix. The participant was updated on their progress throughout the experiment via the software. Finally, the participant filled out the post-experiment questionnaire and the experiment was concluded. This process is illustrated in Figure~\ref{experimentprocess}.

\subsection{Data Processing}

Skin conductance response (SCR) has been shown to be useful in analysis of GSR data \cite{kim2008emotion, kim2004emotion}. We used Ledalab 5 to extract the timing and amplitude of SCR events from the raw GSR data (sampled at 5Hz) using Continuous Decomposition Analysis (CDA) \cite{benedek2010continuous}. Interpolation was performed and the mean, standard deviation, positions of maxima and minima, and number of extrema divided by task duration, were calculated from the SCR amplitude series for each mix \cite{kim2008emotion,daly2015towards}. GSR data of one critical listener was discarded due to poor electrode contact.

We extracted head nod events, head shake events, arousal, expectation, intensity, power and valence from each video clip using the method introduced in \cite{gunes2010dimensional}. Each 20 frames (0.8 sec) of video provided a value for each of these features. Head nod and head shake events are binary values, while the rest of the features are continuous values. We extracted the total head shake and head nod events and took average and standard deviation values for the rest of the features for each video clip.

Intensity values (0-1) of eight AUs, see Table~\ref{auextracted}, were extracted every five frames (0.2 sec) for each video, using the method of \cite{jaiswal2016deep}. We calculated the average and standard deviation values of each AU for each video clip.
\begin{table}[ht!]
\centering
\caption{Extracted Action Units}
\label{auextracted}
\begin{tabular}{@{}ll@{}}
\toprule
\textbf{AU Number} & \textbf{FACS Name} \\ \midrule
AU1                & Inner brow raiser  \\
AU2                & Outer brow raiser  \\
AU4                & Brow lowerer       \\
AU12               & Lip corner puller  \\
AU17               & Chin raiser        \\
AU25               & Lip raiser         \\
AU28               & Lip suck           \\
AU45               & Blink             
\end{tabular}
\end{table}
\vspace{-0.5cm}
\section{Experiments and Results} \label{sec:analysisandresults}
Table~\ref{tab:testconditions} summarises the conditions tested in our experiment. In conditions C1, C2, C5 and C6, we constrained listener type and tested if there was a statistical difference in emotional response ratings and scores based on mix quality. In conditions C3, C4, C7 and C8 we constrained mix quality type and tested if there was a statistical difference in emotional response ratings and scores based on critical listening skills.

We used two types of weightings for ratings and scores, similar to the approaches in \cite{grimm2005evaluation, perez_gonzalez2010real,nicolaou2011continuous}. The audible difference weighting was used in conditions C1 - C4. It weighted participant results by how much they indicated they could hear an audible difference between the high and low quality mix types. The perceived emotional difference weighting was used in conditions C5 - C8, based on how much participants could perceive an emotional difference between the high and low quality mixes. Weights were calculated based on each participant's response to questions asked in the Post-Experiment questionnaire. Each participant indicated on a Likert scale how much they could perceive an audible difference between the two mixes of each song and to what extent they could perceive an emotional difference between the mixes of each song. Weighting was applied as $W_{R} = \frac{O_{R} D_{X}} {N}$, where $O_{R}$ is the original and $W_{R}$ the weighted result, $D_{X}$ is the Likert value for either perceived audible difference or perceived emotional difference, and $N$ is the number of points used in the Likert scale.

In conditions C1, C2, C5 and C6 we used the Wilcoxon Signed Rank non-parametric statistical test because our data is ordinal and we have the same subjects in both datasets. In conditions C3, C4, C7 and C8 we used the Mann-Whitney U non-parametric statistical test because our data is ordinal and we are comparing the medians of two independent groups. In each table in this section the results shown are p-values from the statistical tests for rejecting the null hypothesis, where the numbers in bold are significant $(p < 0.05)$. We have not used the Bonferroni correction because the method is concerned with the general null hypothesis. In this instance, we are investigating how emotions and reactions vary along the many different dimensions tested \cite{perneger1998s}.

The data used for this analysis can be accessed at 

\begin{table*}[ht!]
\centering
\caption{Different types of conditions tested}
\label{tab:testconditions}
\begin{tabular}{@{}ccccc@{}}
\toprule
\textbf{Condition} & \textbf{Constrained}  & \textbf{Varied}                            & \textbf{Weighting}   & \textbf{Statistical Test} \\ \midrule
C1                 & Critical Listener     & High Quality Mix vs Low Quality Mix  & Audible Difference   & Wilcoxon Sign Rank        \\
C2                 & Non-critical Listener & High Quality Mix vs Low Quality Mix  & Audible Difference   & Wilcoxon Sign Rank        \\
C3                 & High Quality Mix    & Critical Listener vs Non-Critical Listener & Audible Difference   & Mann-Whitney U            \\
C4                 & Low Quality Mix   & Critical Listener vs Non-Critical Listener & Audible Difference   & Mann-Whitney U            \\
C5                 & Critical Listener     & High Quality Mix vs Low Quality Mix  & Emotional Difference & Wilcoxon Sign Rank        \\
C6                 & Non-critical Listener & High Quality Mix vs Low Quality Mix  & Emotional Difference & Wilcoxon Sign Rank        \\
C7                 & High Quality Mix    & Critical Listener vs Non-Critical Listener & Emotional Difference & Mann-Whitney U            \\
C8                 & Low Quality Mix   & Critical Listener vs Non-Critical Listener & Emotional Difference & Mann-Whitney U            \\ \bottomrule
\end{tabular}
\end{table*}

\subsection{GEMS-9}
Table~\ref{GEMS9Results} compared the ratings for each of the GEMS-9 emotional adjectives on a song by song basis for conditions C1 to C4. We have removed any p-values that were not significant in order to make the tables easier to read. There are four statistically significant p-values for C1 in contrast to C2 where  there are no statistically significant p-values. This occurred for two songs and happened for the emotions transcendence, tenderness, joyful activation and tension. We see a lot more significant p-values for C3 and C4 than for C1 and C2. We have 47 significant p-values out of a possible 90 for C3 and 43 significant p-values out of 90 for C4. The most amount of significant p-values occur for the emotions of nostalgia, peacefulness, joyful activation and sadness.

\begin{table*}[ht!]
\centering
\caption{\label{GEMS9Results}GEMS-9 - Audible Difference Weighting for Conditions C1 to C4.}
\begin{tabular}{@{}cccccccccc@{}}
\multicolumn{1}{l|}{C1}                 & \textbf{Wonder} & \textbf{Trans} & \textbf{Power} & \textbf{Tender} & \textbf{Nostal} & \textbf{Peace} & \textbf{Joyful} & \textbf{Sadness} & \textbf{Tension} \\ \midrule
\multicolumn{1}{l|}{\textbf{S4}}  &            & \textbf{0.031}         &           &               &               &                 & \textbf{0.031}             &             &            \\

\multicolumn{1}{l|}{\textbf{S7}}  &           &                   &           & \textbf{0.031}      &               &                  &                     &           & \textbf{0.031}   \\
                                      &                 &                        &                &                     &                    &                       &                            &                  &                  \\
\multicolumn{1}{l|}{C2}        & \textbf{Wond} & \textbf{Trans} & \textbf{Power} & \textbf{Tender} & \textbf{Nostal} & \textbf{Peace} & \textbf{Joyful} & \textbf{Sadness} & \textbf{Tension} \\ \midrule

                                      &                 &                        &                &                     &                    &                       &                            &                  &                  \\
\multicolumn{1}{l|}{C3}                 & \textbf{Wonder} & \textbf{Trans} & \textbf{Power} & \textbf{Tender} & \textbf{Nostal} & \textbf{Peace} & \textbf{Joyful} & \textbf{Sadness} & \textbf{Tension} \\ \midrule
\multicolumn{1}{l|}{\textbf{S1}}  &         & \textbf{0.030} & \textbf{0.042} &         &          & \textbf{0.014} & \textbf{0.043} & \textbf{0.011} & \textbf{0.023} \\
\multicolumn{1}{l|}{\textbf{S2}}  &          &          & \textbf{0.039} & \textbf{0.028} &         & \textbf{0.007} &          & \textbf{0.034} &         \\
\multicolumn{1}{l|}{\textbf{S3}}  &          &           &           &           & \textbf{0.024} & \textbf{0.005} &          & \textbf{0.041} &          \\
\multicolumn{1}{l|}{\textbf{S4}}  & \textbf{0.022} &           &           & \textbf{0.018} & \textbf{0.038} & \textbf{0.028} & \textbf{0.007} & \textbf{0.027} &         \\
\multicolumn{1}{l|}{\textbf{S5}}  & \textbf{0.042} &          &          &           & \textbf{0.031} &          &           & \textbf{0.035} &          \\
\multicolumn{1}{l|}{\textbf{S6}}  &           &           & \textbf{0.039} &           & \textbf{0.028} & \textbf{0.041} & \textbf{0.014} &          &         \\
\multicolumn{1}{l|}{\textbf{S7}}  &           &          &           &          &          & \textbf{0.006} & \textbf{0.038} & \textbf{0.038} &          \\
\multicolumn{1}{l|}{\textbf{S8}}  & \textbf{0.035} & \textbf{0.036} & \textbf{0.038} & \textbf{0.031} & \textbf{0.013} &           & \textbf{0.027} &          &          \\
\multicolumn{1}{l|}{\textbf{S9}}  & \textbf{0.027} &         & \textbf{0.043} &         & \textbf{0.030} & \textbf{0.008} & \textbf{0.035} & \textbf{0.042} & \textbf{0.017} \\
\multicolumn{1}{l|}{\textbf{S10}} & \textbf{0.017} &          &          & \textbf{0.022} & \textbf{0.031} & \textbf{0.020} & \textbf{0.027} &          &  \\
                         &                &                &                &                &                &                &                &                &                \\
\multicolumn{1}{l|}{C4}                 & \textbf{Wonder} & \textbf{Trans} & \textbf{Power} & \textbf{Tender} & \textbf{Nostal} & \textbf{Peace} & \textbf{Joyful} & \textbf{Sadness} & \textbf{Tension} \\ \midrule
\multicolumn{1}{l|}{\textbf{S1}}  & \textbf{0.010} & \textbf{0.033} &        &           & \textbf{0.029} & \textbf{0.006} &          & \textbf{0.025} & \textbf{0.049} \\
\multicolumn{1}{l|}{\textbf{S2}}  &          &          & \textbf{0.011} &          &          &           & \textbf{0.023} & \textbf{0.009} &           \\
\multicolumn{1}{l|}{\textbf{S3}}  &         &          & \textbf{0.028} &          & \textbf{0.014} & \textbf{0.005} & \textbf{0.026} &          &          \\
\multicolumn{1}{l|}{\textbf{S4}}  &         &           &           &         &           & \textbf{0.039} &           &          &        \\
\multicolumn{1}{l|}{\textbf{S5}}  & \textbf{0.042} &          &           &          &           & \textbf{0.024} &          &           &          \\
\multicolumn{1}{l|}{\textbf{S6}}  & \textbf{0.034} &          &           &          & \textbf{0.010} & \textbf{0.018} &           & \textbf{0.028} & \textbf{0.020} \\
\multicolumn{1}{l|}{\textbf{S7}}  &         &          &           & \textbf{0.020} & \textbf{0.034} & \textbf{0.004} & \textbf{0.023} &         &           \\
\multicolumn{1}{l|}{\textbf{S8}}  & \textbf{0.017} & \textbf{0.015}          & \textbf{0.045} & \textbf{0.021} & \textbf{0.007} & \textbf{0.006} &          &         &          \\
\multicolumn{1}{l|}{\textbf{S9}}  &          &          & \textbf{0.049} &          & \textbf{0.018} &           &           & \textbf{0.039} & \textbf{0.031} \\
\multicolumn{1}{l|}{\textbf{S10}} & \textbf{0.004} & \textbf{0.016} & \textbf{0.041} & \textbf{0.006} & \textbf{0.011} & \textbf{0.008} & \textbf{0.007} & \textbf{0.032} &  
\end{tabular}
\end{table*}
\subsection{Arousal-Valence-Tension}
Table~\ref{tbl:AVTResults} compares the ratings for Arousal-Valence-Tension dimensions on a song by song basis for Conditions C1 to C4. For C1, there are four statistically significant p-values for arousal, two for valence, and two for tension. This is in contrast to C2 where there is one significant p-value for arousal and one for valence. The significant p-values for C1 are related to six songs in contrast to C2 where they are only related to one song. For both C3 and C4, there are six significant p-values for arousal, all ten for valence and four for tension. p-values for both are similar in terms of distribution over the dimensions, but they differ by song. 

\begin{table}[ht!]
\centering
{\label{tbl:AVTResults}Arousal-Valence-Tension - Audible Difference Weighting for Conditions C1 to C4.}{
\begin{tabular}{ccccccccc}
\multicolumn{1}{l|}{C1}          & \textbf{A}     & \textbf{V}     & \textbf{T}     &  & \multicolumn{1}{l|}{C3}        & \textbf{A}     & \textbf{V}     & \textbf{T}     \\ \cline{1-4} \cline{6-9} 
\multicolumn{1}{l|}{\textbf{S1}}  &          &           &           &  & \multicolumn{1}{l|}{\textbf{S1}}  & \textbf{0.019} & \textbf{0.013} & \textbf{0.045} \\
\multicolumn{1}{l|}{\textbf{S2}}  &         &          &           &  & \multicolumn{1}{l|}{\textbf{S2}}  &           & \textbf{0.011} &        \\
\multicolumn{1}{l|}{\textbf{S3}}  & \textbf{0.021} &         &          &  & \multicolumn{1}{l|}{\textbf{S3}}  &          & \textbf{0.004} & \textbf{0.021} \\
\multicolumn{1}{l|}{\textbf{S4}}  & \textbf{0.002} &         &         &  & \multicolumn{1}{l|}{\textbf{S4}}  &           & \textbf{0.018} &          \\
\multicolumn{1}{l|}{\textbf{S5}}  &           &          &           &  & \multicolumn{1}{l|}{\textbf{S5}}  & \textbf{0.008} & \textbf{0.017} &           \\
\multicolumn{1}{l|}{\textbf{S6}}  &          &          &          &  & \multicolumn{1}{l|}{\textbf{S6}}  & \textbf{0.021} & \textbf{0.009} & \textbf{0.049} \\
\multicolumn{1}{l|}{\textbf{S7}}  & \textbf{0.039} &          &          &  & \multicolumn{1}{l|}{\textbf{S7}}  &          & \textbf{0.002} &          \\
\multicolumn{1}{l|}{\textbf{S8}}  &          & \textbf{0.035} &          &  & \multicolumn{1}{l|}{\textbf{S8}}  & \textbf{0.008} & \textbf{0.006} & \textbf{0.038} \\
\multicolumn{1}{l|}{\textbf{S9}}  & \textbf{0.027} &           & \textbf{0.016} &  & \multicolumn{1}{l|}{\textbf{S9}}  & \textbf{0.009} & \textbf{0.002} &           \\
\multicolumn{1}{l|}{\textbf{S10}} &           & \textbf{0.016} & \textbf{0.031} &  & \multicolumn{1}{l|}{\textbf{S10}} & \textbf{0.019} & \textbf{0.004} &          \\
                                  &                &                &                &  &                                   &                &                &                \\
\multicolumn{1}{l|}{C2}         & \textbf{A}     & \textbf{V}     & \textbf{T}     &  & \multicolumn{1}{l|}{C4}         & \textbf{A}     & \textbf{V}     & \textbf{T}     \\ \cline{1-4} \cline{6-9} 
\multicolumn{1}{l|}{\textbf{S1}}  &          &           &           &  & \multicolumn{1}{l|}{\textbf{S1}}  & \textbf{0.026} & \textbf{0.011} &          \\
\multicolumn{1}{l|}{\textbf{S2}}  & \textbf{0.047} & \textbf{0.039} &          &  & \multicolumn{1}{l|}{\textbf{S2}}  & \textbf{0.007} & \textbf{0.005} &           \\
\multicolumn{1}{l|}{\textbf{S3}}  &          &          &          &  & \multicolumn{1}{l|}{\textbf{S3}}  & \textbf{0.038} & \textbf{0.006} &          \\
\multicolumn{1}{l|}{\textbf{S4}}  &         &          &           &  & \multicolumn{1}{l|}{\textbf{S4}}  &          & \textbf{0.014} &          \\
\multicolumn{1}{l|}{\textbf{S5}}  &          &          &         &  & \multicolumn{1}{l|}{\textbf{S5}}  & \textbf{0.004} & \textbf{0.010} & \textbf{0.010} \\
\multicolumn{1}{l|}{\textbf{S6}}  &          &           &          &  & \multicolumn{1}{l|}{\textbf{S6}}  &          & \textbf{0.005} & \textbf{0.026} \\
\multicolumn{1}{l|}{\textbf{S7}}  &         &          &           &  & \multicolumn{1}{l|}{\textbf{S7}}  &          & \textbf{0.006} &          \\
\multicolumn{1}{l|}{\textbf{S8}}  &           &           &          &  & \multicolumn{1}{l|}{\textbf{S8}}  & \textbf{0.011} & \textbf{0.021} & \textbf{0.041} \\
\multicolumn{1}{l|}{\textbf{S9}}  &          &           &          &  & \multicolumn{1}{l|}{\textbf{S9}}  & \textbf{0.007} & \textbf{0.015} & \textbf{0.028} \\
\multicolumn{1}{l|}{\textbf{S10}} &          &          &           &  & \multicolumn{1}{l|}{\textbf{S10}} &         & \textbf{0.013} &         
\end{tabular}}
\end{table}

\subsection{GSR}
We compared the mean, standard deviation, positions of maxima and minima and frequency of event values for each participant's GSR data on a song by song basis. However, since there were few significant p-values we did not present the results in a table. This was also the only part of the experiment where we tested conditions C1 to C4 as well as conditions C5 to C8, as it was the only time these conditions gave a noticeable amount of significant p-values.

When we tested C1 and C2, there were only 3 out of 50 statistically significant p-values for critical listeners and 3 out of 50 statistically significant p-values for non-critical listeners. Similar results occurred when we tested conditions C5 and C6. C3 gave 5 out of 50 statistically significant p-values for two songs, and there were 4 out of 50 for C4. When we tested condition C7, there were 9 out of 50 statistically significant p-values. This is in contrast to C8 where there were 2 out of 50 statistically significant p-values.  

\subsection{Head Nod and Shake}
We compared Head Nod and Shake scores on a song by song basis. There were no statistically significant p-values for condition C1, and only 2 out 70 p-values for C2 were statistically significant. The results for conditions C3 and C4 are summarised in Table~\ref{HeadNodShakeCriticalBestNonCriticalBest-CriticalWorstNonCriticalWorst}. For C3, we have 31 significant p-values out of a possible 70. The most amount of significant p-values occurred for shake, expectation and power. C4 gave 35 significant p-values out of 70. The largest amount of significant p-values occur for shake, arousal and power.

\begin{table*}[ht!]
\centering
\caption{\label{HeadNodShakeCriticalBestNonCriticalBest-CriticalWorstNonCriticalWorst}Head Nod and Shake - Audible Difference Weighting for Conditions C3 and C4.}
\begin{tabular}{@{}cccccccc@{}}
\multicolumn{1}{l|}{C3}             & \textbf{Nod}   & \textbf{Shake} & \textbf{Arousal} & \textbf{Expectation} & \textbf{Intensity} & \textbf{Power} & \textbf{Valence} \\ \midrule
\multicolumn{1}{l|}{\textbf{S1}}  & \textbf{0.023} & \textbf{0.041} & \textbf{0.006}   & \textbf{0.006}      & \textbf{0.006}     &         &           \\
\multicolumn{1}{l|}{\textbf{S2}}  &         & \textbf{0.017} &            &             &            & \textbf{0.034} &             \\
\multicolumn{1}{l|}{\textbf{S3}}  &          & \textbf{0.002} & \textbf{0.009}   & \textbf{0.004}      & \textbf{0.017}     & \textbf{0.000} &           \\
\multicolumn{1}{l|}{\textbf{S4}}  &          &          &            &               &               & \textbf{0.009} & \textbf{0.002}   \\
\multicolumn{1}{l|}{\textbf{S5}}  & \textbf{0.026} &          & \textbf{0.006}   &              & \textbf{0.006}     &           &           \\
\multicolumn{1}{l|}{\textbf{S6}}  &         & \textbf{0.013} &             & \textbf{0.026}      &              & \textbf{0.038} &           \\
\multicolumn{1}{l|}{\textbf{S7}}  &          &           &            & \textbf{0.014}      &               &          &             \\
\multicolumn{1}{l|}{\textbf{S8}}  &          & \textbf{0.011} & \textbf{0.021}   & \textbf{0.009}      &              & \textbf{0.031} &            \\
\multicolumn{1}{l|}{\textbf{S9}}  &         & \textbf{0.005} & \textbf{0.001}   & \textbf{0.001}      & \textbf{0.002}     & \textbf{0.003} &          \\
\multicolumn{1}{l|}{\textbf{S10}} &          &          &            & \textbf{0.002}      &               &          &            \\
                                  &                &                &                  &                     &                    &                &                  \\
\multicolumn{1}{l|}{C4}             & \textbf{Nod}   & \textbf{Shake} & \textbf{Arousal} & \textbf{Expectation} & \textbf{Intensity} & \textbf{Power} & \textbf{Valence} \\ \midrule
\multicolumn{1}{l|}{\textbf{S1}}  &          & \textbf{0.005} &            &                &               & \textbf{0.026} &             \\
\multicolumn{1}{l|}{\textbf{S2}}  & \textbf{0.006} &          & \textbf{0.010}   & \textbf{0.009}      & \textbf{0.010}     &           &            \\
\multicolumn{1}{l|}{\textbf{S3}}  & \textbf{0.028} &          &             &                &               & \textbf{0.014} &            \\
\multicolumn{1}{l|}{\textbf{S4}}  &        &          &            &              &              & \textbf{0.045} & \textbf{0.007}   \\
\multicolumn{1}{l|}{\textbf{S5}}  & \textbf{0.034} & \textbf{0.036} & \textbf{0.017}   &              &              &           &           \\
\multicolumn{1}{l|}{\textbf{S6}}  &          & \textbf{0.007} & \textbf{0.038}   & \textbf{0.031}      & \textbf{0.045}     & \textbf{0.031} &           \\
\multicolumn{1}{l|}{\textbf{S7}}  & \textbf{0.005} &          & \textbf{0.007}   & \textbf{0.011}      &              & \textbf{0.005} &             \\
\multicolumn{1}{l|}{\textbf{S8}}  &          & \textbf{0.001} & \textbf{0.017}   & \textbf{0.000}      & \textbf{0.021}     & \textbf{0.004} &            \\
\multicolumn{1}{l|}{\textbf{S9}}  & \textbf{0.017} & \textbf{0.017} & \textbf{0.021}   &               &               & \textbf{0.034} &            \\
\multicolumn{1}{l|}{\textbf{S10}} &           & \textbf{0.006} & \textbf{0.028}   & \textbf{0.023}      & \textbf{0.013}     &           &           
\end{tabular}
\end{table*}

\subsection{Facial Action Units}
We compared the standard deviation for each participant's Facial Action Unit scores on a song by song basis. We saw 3 out of 80 statistically significant p-values for condition C1, whereas C2 gave 7 out of 80 statistically significant p-values. Results for conditions C3 and C4 are summarised in Table~\ref{FACSCriticalBestNonCriticalBest-CriticalWorstNonCriticalWorst}. There were 23 significant p-values out of a possible 80, mainly for AU1, AU4 and AU45. For condition C4, we have 20 significant p-values out of 80, mostly from AU4 and AU45.

We also examined which AUs had the highest intensity throughout the experiment. We checked every mix that each participant listened to, to see if any of their average AU intensities was $>= 0.5$. If the average AU intensity was $>= 0.5$ we marked the AU for that particular mix with a 1, otherwise a 0. We summarised the results as a percentage of all the mixes listened to for critical listeners and non-critical listeners in Table~\ref{tab:auintensity}. AU1 and AU4 gave the greatest amount of average AU intensities $>= 0.5$. The results for AU12 and AU17 were omitted since all the results were 0. Critical listeners experienced a greater number of average AU intensities $>= 0.5$ than non-critical listeners for all AUs except AU28. However, the difference in the case of AU28 is 0.005, which is negligible. 

\begin{table*}[ht!]
\centering
\caption{\label{FACSCriticalBestNonCriticalBest-CriticalWorstNonCriticalWorst}FACS - Audible Difference Weighting for Conditions C3 and C4.}
\begin{tabular}{ccccccccc}
\multicolumn{1}{l|}{C3}              & \textbf{AU1}   & \textbf{AU2} & \textbf{AU4}   & \textbf{AU12} & \textbf{AU17}  & \textbf{AU25}  & \textbf{AU28} & \textbf{AU45}  \\ \hline
\multicolumn{1}{l|}{\textbf{S1}}  & \textbf{0.011} &       & \textbf{0.021} &        &           &          &         &           \\
\multicolumn{1}{l|}{\textbf{S2}}  & \textbf{0.038} &        & \textbf{0.006} &          &           &           &         &           \\
\multicolumn{1}{l|}{\textbf{S3}}  &           &         & \textbf{0.026} &         &           &           &          & \textbf{0.038} \\
\multicolumn{1}{l|}{\textbf{S4}}  & \textbf{0.026} &        & \textbf{0.014} &          &         & \textbf{0.038} &          &           \\
\multicolumn{1}{l|}{\textbf{S5}}  & \textbf{0.045} &        &          &          & \textbf{0.007} &           &          &          \\
\multicolumn{1}{l|}{\textbf{S6}}  &           &        & \textbf{0.004} &         &         &         &          & \textbf{0.045} \\
\multicolumn{1}{l|}{\textbf{S7}}  & \textbf{0.038} &        & \textbf{0.006} &        & \textbf{0.011} &         &         & \textbf{0.031} \\
\multicolumn{1}{l|}{\textbf{S8}}  & \textbf{0.038} &       & \textbf{0.004} &         &          &          &         & \textbf{0.026} \\
\multicolumn{1}{l|}{\textbf{S9}}  &          &       & \textbf{0.038} &          &         &           &         & \textbf{0.014} \\
\multicolumn{1}{l|}{\textbf{S10}} &          &         &           &          &          &         &          & \textbf{0.021} \\
                                      &                &              &                &               &                &                &               &                \\
\multicolumn{1}{l|}{C4}             & \textbf{AU1}   & \textbf{AU2} & \textbf{AU4}   & \textbf{AU12} & \textbf{AU17}  & \textbf{AU25}  & \textbf{AU28} & \textbf{AU45}  \\ \hline
\multicolumn{1}{l|}{\textbf{S1}}  &          &         & \textbf{0.009} &          &           &           &          & \textbf{0.045} \\
\multicolumn{1}{l|}{\textbf{S2}}  & \textbf{0.031} &        & \textbf{0.045} &          &           &          & 
& \textbf{0.045} \\
\multicolumn{1}{l|}{\textbf{S3}}  & \textbf{0.003} &         & \textbf{0.007} &          &           &          &          &           \\
\multicolumn{1}{l|}{\textbf{S4}}  &         &         & \textbf{0.009} &          &           &          &          & \textbf{0.038} \\
\multicolumn{1}{l|}{\textbf{S5}}  &          &         &           &          &         &           &          & \textbf{0.006} \\
\multicolumn{1}{l|}{\textbf{S6}}  &          &         & \textbf{0.002} &          & \textbf{0.031} &           &          & \textbf{0.011} \\
\multicolumn{1}{l|}{\textbf{S7}}  &          &         & \textbf{0.021} &          &         &           &        &           \\
\multicolumn{1}{l|}{\textbf{S8}}  & \textbf{0.045} &        & \textbf{0.014} &       &           &          &         & \textbf{0.011} \\
\multicolumn{1}{l|}{\textbf{S9}}  &          &       & \textbf{0.009} &          &          &           &          &           \\
\multicolumn{1}{l|}{\textbf{S10}} &          &         & \textbf{0.006} &         &           &          &          & \textbf{0.026}
\end{tabular}
\end{table*}

\begin{table*}[ht!]
\centering
{\label{tab:auintensity}Percentage of mixes where average AU intensity was $>=$ 0.5. (i) Non-critical listeners (ii) Critical listeners}{
\begin{tabular}{c|cccccccc|cccccc}
(i)               & \textbf{AU1} & \textbf{AU2} & \textbf{AU4} & \textbf{AU25} & \textbf{AU28} & \textbf{AU45} &  & (ii)              & \textbf{AU1} & \textbf{AU2} & \textbf{AU4} & \textbf{AU25} & \textbf{AU28} & \textbf{AU45} \\ \cline{1-7} \cline{9-15} 
\textbf{A}        & 0.9          &             &             &              &              &              &  & \textbf{K}        & 1            &             & 1            &              &              &              \\
\textbf{B}        & 0.85         &             & 0.85         &              &              &              &  & \textbf{L}        & 0.95         & 0.05         & 0.25         & 0.25          &              & 0.1           \\
\textbf{C}        & 0.55         &             & 0.7          &              &              &              &  & \textbf{M}        & 0.1          &             & 0.95         &              &              & 0.2           \\
\textbf{D}        &             &             &             &              &              &              &  & \textbf{N}        & 0.55         &             & 0.35         &              &              &             \\
\textbf{E}        &             & 0.05         & 0.95         &              &              & 0.05          &  & \textbf{O}        &             &             & 1            &              &              &              \\
\textbf{F}        &             &             & 0.75         & 0.05          &              &              &  & \textbf{P}        & 0.9          &             & 1            &              &              &              \\
\textbf{G}        & 0.25         &             & 0.55         &              &              &              &  & \textbf{Q}        & 0.45         &             & 1            &              &              &              \\
\textbf{H}        & 1            &             & 0.75         &              & 0.05          &              &  & \textbf{R}        & 0.2          &             & 0.35         & 0.1           &              & 0.15          \\
\textbf{I}        & 0.75         &             & 0.7          & 0.05          &              &              &  & \textbf{S}        & 0.75         & 0.45         & 1            &              &              &              \\
\textbf{J}        &             &             & 0.85         &              &              &              &  & \textbf{T}        & 0.8          &             & 0.05         &              &              &              \\ \cline{1-7} \cline{9-15} 
\textbf{Total \%} & 0.43         & 0.005        & 0.61         & 0.01          & 0.005         & 0.005         &  & \textbf{Total \%} & 0.57         & 0.05         & 0.695        & 0.035         &   0           & 0.045        
\end{tabular}}
\end{table*}
\section{Discussion} \label{sec:discussion}

\subsection{Findings}
\subsubsection{GEMS-9}
With GEMS-9 we investigated if there was a significant difference in the distribution of induced emotions of each listener type. Table \ref{GEMS9Results} results indicate that the critical listeners were the only group where there was significant differences in the distribution of induced emotions between the two mix types. This suggests that our hypothesis is true. However, since there are so few p-values in comparison to the amount of tests we can not draw a strong conclusion from this. 

Results also indicate that high quality mixes had a greater significant difference on the distribution of induced emotions between the two listener types. These results support our hypothesis, in that the high quality mix had more of an impact emotionally on one listener type over the other. They also imply that there was a greater difference in the indicated levels of joyful activation and sadness between critical and non-critical listeners for the high quality mixes (C3). Joyful activation and sadness would be synonymous with the positive and negative valence, implying that the quality of the mix may have an impact on how happy or sad a critical listener may feel.

\subsubsection{Arousal-Valence-Tension}
We investigated if there was a significant difference in the distribution of emotions perceived by each listener type along Arousal-Valence-Tension dimensions. Table~\ref{tbl:AVTResults} indicates that for critical listeners there are more examples of where there are significant differences in the distribution of perceived emotions, especially with respect to arousal. This was the only time a noticeable difference in the amount of significant p-values occurred when we compared the critical listener's high quality mixes to critical listener's low quality mixes. This also occurred in the case of non-critical listeners (C2), but to a lesser extent. These results support our hypothesis, in that critical listeners were able to perceive an emotional difference between the two mixes much more so than non-critical listeners and this was mostly with respect to arousal and tension.
 
Table~\ref{tbl:AVTResults} showed a lot of significant p-values for Conditions C3 and C4 in comparison to C1 and C2. Interestingly, we have the same amount of significant values in each dimension for both conditions C3 and C4. This implies that there are the same amount of significant differences in the distribution of emotions for both listener types due to mix quality, but it varies by song. The two listener types are perceiving different levels of arousal and tension, but on different songs. However, this may have something to do with the participant's genre preference. These results are similar to those seen in Table~\ref{GEMS9Results} (iii) and (iv), in the respect that joyful activation corresponds to positive valence and sadness corresponds to negative valence.

\subsubsection{GSR}
Overall GSR gave largely inconclusive results except when we examined response of critical and non-critical listener's to high quality mixes (C3, C7). There is also a trend when we compare the results for C3 and C7, against the results for critical and non-critical listeners' low quality mixes (C4, C8). There are more significant results when we do this comparison as opposed to comparing responses of critical listener's to high and low quality mixes (C1, C5), against responses of non-critical listener's to high and low quality mixes (C2, C6). We also saw this for GEMS-9 and Arousal-Valence-Tension. Thus testing critical versus non-critical listener responses to high versus low quality mixes supported our hypothesis.

\subsubsection{Head Nod and Shake}

Head nod/shake results proved to be conclusive and supported our hypothesis. The difference in nodding is far more apparent for low quality mixes (C4) than high quality mixes (C3).  Notably, on low quality mixes, non-critical listeners nodded their heads more than critical listeners. This could mean that non-critical listeners  might enjoy the mix regardless of mix quality. We also see something similar for arousal and power where there are slightly more significant p-values for the low quality mixes than for the high quality mixes.

Power, expectation and arousal seem to be divisive features when comparing the types of listeners. Power is based on the sense of control, expectation on the degree of anticipation and arousal on the degree of excitement or apathy \cite{gunes2010dimensional}. These are features based on tracking emotional cues when conversing with someone, so it is interesting to see them having such an effect during music listening. Having examined the participant's videos we found that since they were sitting in a chair that could rotate, they sometimes moved the chair in time with the music. The classifier detected this as a head shake, which would normally be viewed as a negative response \cite{tom1991role}, but in this case it could indicate that the participant is engaged with the music and most likely enjoying it. It is also worth noting that music is very cultural and certain individuals might react differently than others with respect to head nods and shakes.

\subsubsection{Facial Action Units}
Results indicated that the high quality mixes had a greater effect than low quality mixes on the distribution of AU1 and AU4 between the two listener types. AU1 corresponds to inner brow raiser and AU4 corresponds to brow lowering, so this is similar to research on Facial EMG and music, where the brow is associated with the processing of negative events \cite{schwartz1980facial, witvliet2007play}. AU45 corresponds to blinking. There is one more significant AU45 result for condition C4 than there is condition C3, which might imply that there is a difference in intensity of blinking for critical and non-critical listeners.
    
The percentage total of average AU intensities $>= 0.5$ for AU45 is small, but provided a large amount of significant p-values in Table~\ref{FACSCriticalBestNonCriticalBest-CriticalWorstNonCriticalWorst}. This suggests that the differences in blink intensity between listener type may have been very subtle. 

\begin{figure}[ht!]
\centering
\includegraphics[scale=	0.12]{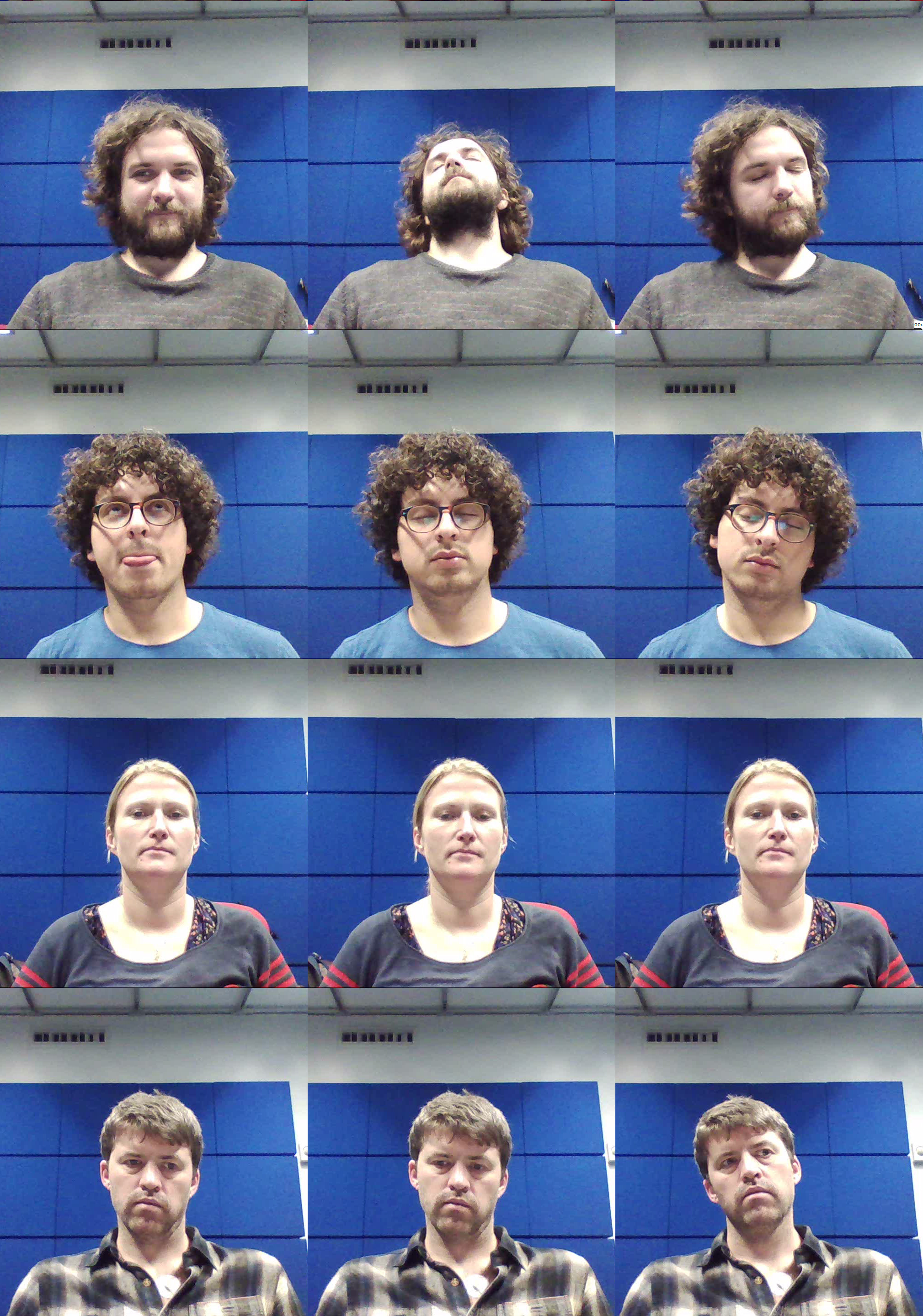}
\caption{\label{fig:filmstrip} Still images of four participants from the videos made during the experiment. Top two rows are critical listeners and the bottom two are non-critical listeners.}
\end{figure}

This is the first experiment of its kind that has looked at automatic facial expression recognition and tracking head nod/shakes in a music production quality context. By inspecting the videos we found that some participants were much more expressive in their face than others or might be a lot more inclined to nod and shake their head than use facial expressions. Some critical listeners gazed left or right of the camera, closed their eyes while listening for a prolonged duration, placed their hand under their chin, looked down, looked up, moved their head back and forth, tilted their head or sucked their lip. For non-critical listeners, there were not as many AU's activated, except in one case where the participant was looking away, moving their body on the chair left and right, moving their head back and forth and moving their head left and right. Some stills from the videos can be seen in Figure~\ref{fig:filmstrip}, where the top two participants are critical listeners and the bottom two are non-critical listeners.

\begin{figure}[ht!]
\centering
\includegraphics[scale=	0.53]{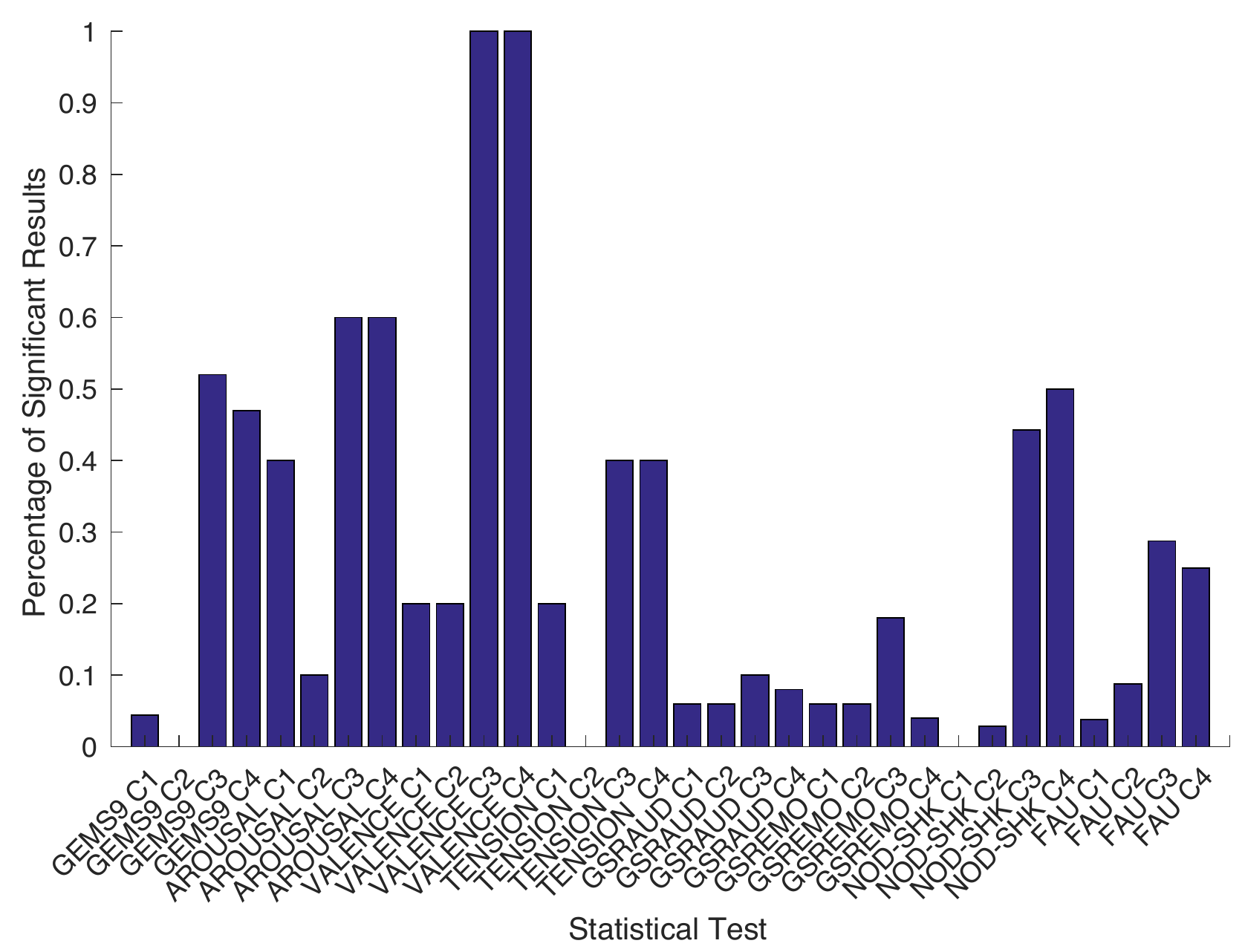}
\caption{\label{fig:statresutls} The percentage of significant results for each statistical test performed for each condition. The highest percentage of significant results occurred for GEMS9 (Felt emotion), Arousal-Valence-Tension (Perceived emotion), Head Nod/Shake and Facial Action Units.}
\end{figure}

\subsection{Measures}

Self-report measures proved to be the most revealing when comparing mixes and when comparing listener types. We expected the GSR results to be more telling, but found them to be mostly inconclusive. This might have been due to noise in the data as a result of poor electrode contact which is similar to what happened in \cite{morgan2015using}.

The values for the AU’s only became interesting when we looked at the standard deviation. This is expected since someone that is more excited by music tends to be more expressive in their face as the music is played. Head nod/shake detection proved to be very interesting when comparing the types of listeners. Non-critical listeners nodded their heads more than critical listeners when listening to the poor quality mix, which was something we decided to analyse based on our initial findings in the pilot study. 

\subsection{Design}

As beneficial as it was to have a pilot study, we learned a lot about experimental design from the main part of the experiment, which could be used to help future studies. One participant reported that most of the emotions that music induces for them comes from the lyrics. They reported that if they disliked the lyrics, then they tended to dislike the song, thus potentially meaning a negative or lack of emotional response. This aspect of music listening may have had an impact on the emotional responses of non-native English speakers. Ten of the participants were non-native speakers and may not have fully understood all lyrics, so this is a confounding variable we had not considered \cite{rentfrow2003re}.

Recent research on perceptual evaluation of high resolution audio found that providing training before conducting perceptual experiments greatly improved the reliability of results \cite{reisshighresaudio2016}. In our experiment we provided two training songs, but this was to become familiar with the experimental interface. However, it could be argued that training would have blurred the distinction between critical and non-critical listeners.

Ideally we would have used songs in the experiment that came from a wider variety of genres. A number of participants were dissatisfied with the songs because they simply did not like the genre. But this was out of our control since we used songs rated in a previous experiment \cite{deman2014d}. We would have also liked to have had a bigger sample size for our experiment, to further generalise the results.

We would also suggest that each participant be made sit on a chair that does not rotate or have wheels. When some participants were enjoying a song they tended to move around, which sometimes caused sensors to become dislodged and rendered the acquired data unusable.

\section{Conclusion} \label{sec:conclusion}
Our exploratory study provides an insight into the relationship between music production quality and musically induced and perceived emotions. We highlighted some of the challenges with working with physiological sensors and conducting listening tests when trying to measure emotional responses in a musical context. We conducted the first experiment of its kind using facial expression analysis and head nod-shake detection in conjunction with a perceptual listening test. 

When we tested to see if critical listeners and non-critical listeners had different emotional responses based on the difference in music production quality, the results were inconclusive for GSR, facial expression and head nod-shake detection. Results strongly agreed with our hypothesis only when we looked at the self-report of perceived emotion.

When we examined just high quality mixes and looked at the difference in emotions of critical and non-critical listeners we found significant p-values in most cases. This was most evident for self-report, head nods/shakes and facial expression. When we examined low quality mixes and looked at the difference in emotions of critical and non-critical listeners we also found a lot of significant p-values, but to a lesser extent than that of the high quality mixes. This was also most evident for self-report, head nods/shakes and facial expression.

The results implied that emotion in a mix, whether induced or perceived, mattered the most to those with critical listening skills, which agrees with our hypothesis. This was most evident from the GEMS-9, Arousal-Valence-Tension, Head Nod/Shake Detection and Facial Action Unit results since they had the most amount of significant p-values. 

If one was to take a cynical view, it could be said that using a more professional and experienced mix engineer to mix a piece of music only really matters to those who have been trained to listen for mix defects, and mix quality has little bearing on the layperson emotionally. This is an important result for audio engineers and specifically in the context of automatic mixing systems. The perceived quality of an automatically generated mix may not be important to those without critical listening skills and it suggests that automatically generated mixes may be good enough for the general public.

\section{Future Work} \label{sec:futurework}
It would be interesting to perform pair-wise ranking between the two mix types, as Likert scales may not be the best tool for affect studies since the values they ask people to rate may mean different things to each participant \cite{yannakakis2011ranking}. However, one argument against pairwise testing is that it is time consuming, e.g. for 10 samples, one might need 10*9/2 comparisons \cite{schatz2012impact, ackerman2009test}.

It would also be interesting to see if we get similar results when non-critical listeners are provided with training before the experiment i.e. trained to spot common mix defects. This would help identify if the trained non-critical listener's exhibited emotions based on what they think is expected of them due to the training. 

We would like to track if a participant is singing along to the music being played, as this could be regarded as a measure of engagement and potential enjoyment of the music. This could be achieved by tracking the Action Units that correspond to the mouth as well as having a microphone near the participant to verify if they were actually singing or not. We would also recommend looking at tracking foot or finger tapping as this is a common form of movement to music \cite{wallin2001origins}. This could be achieved by attaching accelerometers to the participant's feet and placing small piezo contact microphones on their fingertips.

We hope this work will inspire future research. In particular there is a need to use more varied genres of music for evaluation and to see if emotional measures correlate well with low to high level audio features. This could potentially be used in automatic mixing systems such as \cite{hafezi2015autonomous, reiss2011intelligent, ronan15subgrouping,macompression}.

\textbf{Acknowledgements:} The authors would like to thank all the participants of this study and EPSRC UK for funding this research. We would also like to thank Elio Quinton, Dave Moffat and Emmanuel Deruty for providing valuable feedback.

\ifCLASSOPTIONcaptionsoff
  \newpage
\fi



%
\bibliography{references} 

\begin{thebibliography}{10}

\bibitem{case2011mix}
A.~U. Case, {\em Mix smart}.
\newblock Focal Press, 2011.

\bibitem{deman2015perceptual}
B.~De~Man, M.~Boerum, B.~Leonard, R.~King, G.~Massenburg, and J.~D. Reiss,
  ``Perceptual evaluation of music mixing practices,'' in {\em 138th Convention
  of the Audio Engineering Society}, May 2015.

\bibitem{deman2014d}
B.~De~Man, B.~Leonard, R.~King, and J.~D. Reiss, ``An analysis and evaluation
  of audio features for multitrack music mixtures,'' in {\em 15th International
  Society for Music Information Retrieval Conference (ISMIR 2014)}, October
  2014.

\bibitem{ronan2015automatic}
D.~Ronan, D.~Moffat, H.~Gunes, and J.~D. Reiss, ``Automatic subgrouping of
  multitrack audio,'' in {\em Proc. 18th International Conference on Digital
  Audio Effects (DAFx-15)}, 2015.

\bibitem{ronan2015impact}
D.~Ronan, B.~De~Man, H.~Gunes, and J.~D. Reiss, ``The impact of subgrouping
  practices on the perception of multitrack mixes,'' in {\em 139th Convention
  of the Audio Engineering Society}, 2015.

\bibitem{wilson2015perception}
A.~Wilson and B.~M. Fazenda, ``Perception of audio quality in productions of
  popular music,'' {\em Journal of the Audio Engineering Society}, 2015.

\bibitem{pestana2014intelligent}
P.~Pestana and J.~Reiss, ``Intelligent audio production strategies informed by
  best practices,'' in {\em Audio Engineering Society Conference: 53rd
  International Conference: Semantic Audio}, Audio Engineering Society, 2014.

\bibitem{ross2007rest}
A.~Ross, {\em The rest is noise: Listening to the twentieth century}.
\newblock Macmillan, 2007.

\bibitem{gabrielsson2002emotion}
A.~Gabrielsson, ``Emotion perceived and emotion felt: Same or different?,''
  {\em Musicae Scientiae}, vol.~5, no.~1 suppl, pp.~123--147, 2002.

\bibitem{eerola2010comparison}
T.~Eerola and J.~K. Vuoskoski, ``A comparison of the discrete and dimensional
  models of emotion in music,'' {\em Psychology of Music}, 2010.

\bibitem{song2016perceived}
Y.~Song, S.~Dixon, M.~T. Pearce, and A.~R. Halpern, ``Perceived and induced
  emotion responses to popular music,'' {\em Music Perception: An
  Interdisciplinary Journal}, vol.~33, no.~4, pp.~472--492, 2016.

\bibitem{gabrielsson2010role}
A.~Gabrielsson and E.~Lindstr{\"o}m, ``The role of structure in the musical
  expression of emotions,'' {\em Handbook of music and emotion: Theory,
  research, applications}, pp.~367--400, 2010.

\bibitem{juslin2008emotional}
P.~N. Juslin and D.~V{\"a}stfj{\"a}ll, ``Emotional responses to music: The need
  to consider underlying mechanisms,'' {\em Behavioral and brain sciences},
  vol.~31, no.~05, pp.~559--575, 2008.

\bibitem{juslin2010does}
P.~N. Juslin, S.~Liljestr{\"o}m, D.~V{\"a}stfj{\"a}ll, and L.-O. Lundqvist,
  ``How does music evoke emotions? exploring the underlying mechanisms,'' in
  {\em Handbook of music and emotion}, pp.~605--642, Oxford Press, 2010.

\bibitem{juslin2013everyday}
P.~N. Juslin, ``From everyday emotions to aesthetic emotions: towards a unified
  theory of musical emotions,'' {\em Physics of life reviews}, vol.~10, no.~3,
  pp.~235--266, 2013.

\bibitem{juslin2013makes}
P.~N. Juslin, L.~Harmat, and T.~Eerola, ``What makes music emotionally
  significant? exploring the underlying mechanisms,'' {\em Psychology of
  Music}, p.~0305735613484548, 2013.

\bibitem{ekman1992argument}
P.~Ekman, ``An argument for basic emotions,'' {\em Cognition \& emotion},
  vol.~6, no.~3-4, pp.~169--200, 1992.

\bibitem{panksepp1998affective}
J.~Panksepp, {\em Affective neuroscience: The foundations of human and animal
  emotions}.
\newblock Oxford university press, 1998.

\bibitem{eerola2009prediction}
T.~Eerola, O.~Lartillot, and P.~Toiviainen, ``Prediction of multidimensional
  emotional ratings in music from audio using multivariate regression
  models.,'' in {\em 10th International Society for Music Information Retrieval
  Conference (ISMIR 2009)}, pp.~621--626, 2009.

\bibitem{sloboda2001psychological}
J.~A. Sloboda and P.~N. Juslin, ``Psychological perspectives on music and
  emotion.,'' 2001.

\bibitem{russell1980circumplex}
J.~A. Russell, ``A circumplex model of affect.,'' {\em Journal of personality
  and social psychology}, vol.~39, no.~6, p.~1161, 1980.

\bibitem{zentner2008emotions}
M.~Zentner, D.~Grandjean, and K.~R. Scherer, ``Emotions evoked by the sound of
  music: characterization, classification, and measurement.,'' {\em Emotion},
  vol.~8, no.~4, p.~494, 2008.

\bibitem{pearce2015age}
M.~T. Pearce and A.~R. Halpern, ``Age-related patterns in emotions evoked by
  music.,'' 2015.

\bibitem{kreutz2007using}
G.~Kreutz, U.~Ott, D.~Teichmann, P.~Osawa, and D.~Vaitl, ``Using music to
  induce emotions: Influences of musical preference and absorption,'' {\em
  Psychology of music}, 2007.

\bibitem{vieillard2008happy}
S.~Vieillard, I.~Peretz, N.~Gosselin, S.~Khalfa, L.~Gagnon, and B.~Bouchard,
  ``Happy, sad, scary and peaceful musical excerpts for research on emotions,''
  {\em Cognition \& Emotion}, vol.~22, no.~4, pp.~720--752, 2008.

\bibitem{bradley1994measuring}
M.~M. Bradley and P.~J. Lang, ``Measuring emotion: the self-assessment manikin
  and the semantic differential,'' {\em Journal of behavior therapy and
  experimental psychiatry}, vol.~25, no.~1, pp.~49--59, 1994.

\bibitem{izard2007basic}
C.~E. Izard, ``Basic emotions, natural kinds, emotion schemas, and a new
  paradigm,'' {\em Perspectives on psychological science}, vol.~2, no.~3,
  pp.~260--280, 2007.

\bibitem{watson1988development}
D.~Watson, L.~A. Clark, and A.~Tellegen, ``Development and validation of brief
  measures of positive and negative affect: the panas scales.,'' {\em Journal
  of personality and social psychology}, vol.~54, no.~6, p.~1063, 1988.

\bibitem{grewe2007emotions}
O.~Grewe, F.~Nagel, R.~Kopiez, and E.~Altenm{\"u}ller, ``Emotions over time:
  Synchronicity and development of subjective, physiological, and facial
  affective reactions to music.,'' {\em Emotion}, vol.~7, no.~4, p.~774, 2007.

\bibitem{hunter2010music}
P.~G. Hunter and E.~G. Schellenberg, ``Music and emotion,'' in {\em Music
  perception}, pp.~129--164, Springer, 2010.

\bibitem{schubert2010continuous}
E.~Schubert, ``Continuous self-report methods,'' {\em Handbook of music and
  emotion: Theory, research, applications}, vol.~2, pp.~223--253, 2010.

\bibitem{egermann2013probabilistic}
H.~Egermann, M.~T. Pearce, G.~A. Wiggins, and S.~McAdams, ``Probabilistic
  models of expectation violation predict psychophysiological emotional
  responses to live concert music,'' {\em Cognitive, Affective, \& Behavioral
  Neuroscience}, vol.~13, no.~3, pp.~533--553, 2013.

\bibitem{morgan2015using}
E.~Morgan, H.~Gunes, and N.~Bryan-Kinns, ``Using affective and behavioural
  sensors to explore aspects of collaborative music making,'' {\em
  International Journal of Human-Computer Studies}, vol.~82, pp.~31--47, 2015.

\bibitem{hodges2010psychophysiological}
D.~Hodges, ``Psychophysiological measures,'' {\em Handbook of music and
  emotion}, pp.~279--312, 2010.

\bibitem{schwartz1980facial}
G.~E. Schwartz, S.-L. Brown, and G.~L. Ahern, ``Facial muscle patterning and
  subjective experience during affective imagery: Sex differences,'' {\em
  Psychophysiology}, vol.~17, no.~1, pp.~75--82, 1980.

\bibitem{witvliet2007play}
C.~V. Witvliet and S.~R. Vrana, ``Play it again sam: Repeated exposure to
  emotionally evocative music polarises liking and smiling responses, and
  influences other affective reports, facial emg, and heart rate,'' {\em
  Cognition and Emotion}, vol.~21, no.~1, pp.~3--25, 2007.

\bibitem{andreassi2013psychophysiology}
J.~L. Andreassi, {\em Psychophysiology: Human behavior \& physiological
  response}.
\newblock Psychology Press, 2013.

\bibitem{lundqvist2008emotional}
L.-O. Lundqvist, F.~Carlsson, P.~Hilmersson, and P.~Juslin, ``Emotional
  responses to music: experience, expression, and physiology,'' {\em Psychology
  of Music}, 2008.

\bibitem{roy2009modulation}
M.~Roy, J.-P. Mailhot, N.~Gosselin, S.~Paquette, and I.~Peretz, ``Modulation of
  the startle reflex by pleasant and unpleasant music,'' {\em International
  Journal of Psychophysiology}, vol.~71, no.~1, pp.~37--42, 2009.

\bibitem{hager2002facial}
J.~C. Hager, P.~Ekman, and W.~V. Friesen, ``Facial action coding system,'' {\em
  Salt Lake City, UT: A Human Face}, 2002.

\bibitem{weisgerber2015facial}
A.~Weisgerber, N.~Vermeulen, I.~Peretz, S.~Samson, P.~Philippot, P.~Maurage,
  D.~Catherine De~Graeuwe, A.~De~Jaegere, B.~Delatte, B.~Gillain, {\em et~al.},
  ``Facial, vocal and musical emotion recognition is altered in paranoid
  schizophrenic patients,'' {\em Psychiatry research}, 2015.

\bibitem{silvey2012role}
B.~A. Silvey, ``The role of conductor facial expression in students evaluation
  of ensemble expressivity,'' {\em Journal of Research in Music Education},
  p.~0022429412462580, 2012.

\bibitem{wallin2001origins}
N.~L. Wallin and B.~Merker, {\em The origins of music}.
\newblock MIT press, 2001.

\bibitem{sedlmeier2011music}
P.~Sedlmeier, O.~Weigelt, and E.~Walther, ``Music is in the muscle: How
  embodied cognition may influence music preferences,'' {\em Music Perception:
  An Interdisciplinary Journal}, vol.~28, no.~3, pp.~297--306, 2011.

\bibitem{tom1991role}
G.~Tom, P.~Pettersen, T.~Lau, T.~Burton, and J.~Cook, ``The role of overt head
  movement in the formation of affect,'' {\em Basic and Applied Social
  Psychology}, vol.~12, no.~3, pp.~281--289, 1991.

\bibitem{wells1980effects}
G.~L. Wells and R.~E. Petty, ``The effects of over head movements on
  persuasion: Compatibility and incompatibility of responses,'' {\em Basic and
  Applied Social Psychology}, vol.~1, no.~3, pp.~219--230, 1980.

\bibitem{deman2014omtb}
B.~De~Man, M.~Mora-Mcginity, G.~Fazekas, and J.~D. Reiss, ``{T}he {O}pen
  {M}ultitrack {T}estbed,'' in {\em 137th Convention of the Audio Engineering
  Society}, October 2014.

\bibitem{itu2011itu}
R.~ITU-R, ``Itu-r bs. 1770-2, algorithms to measure audio programme loudness
  and true-peak audio level,'' {\em International Telecommunications Union,
  Geneva}, 2011.

\bibitem{kim2008emotion}
J.~Kim and E.~Andr{\'e}, ``Emotion recognition based on physiological changes
  in music listening,'' {\em IEEE transactions on pattern analysis and machine
  intelligence}, vol.~30, no.~12, pp.~2067--2083, 2008.

\bibitem{kim2004emotion}
K.~H. Kim, S.~Bang, and S.~Kim, ``Emotion recognition system using short-term
  monitoring of physiological signals,'' {\em Medical and biological
  engineering and computing}, vol.~42, no.~3, pp.~419--427, 2004.

\bibitem{benedek2010continuous}
M.~Benedek and C.~Kaernbach, ``A continuous measure of phasic electrodermal
  activity,'' {\em Journal of neuroscience methods}, vol.~190, no.~1,
  pp.~80--91, 2010.

\bibitem{daly2015towards}
I.~Daly, A.~Malik, J.~Weaver, F.~Hwang, S.~J. Nasuto, D.~Williams, A.~Kirke,
  and E.~Miranda, ``Towards human-computer music interaction: Evaluation of an
  affectively-driven music generator via galvanic skin response measures,'' in
  {\em Computer Science and Electronic Engineering Conference (CEEC), 2015
  7th}, pp.~87--92, IEEE, 2015.

\bibitem{gunes2010dimensional}
H.~Gunes and M.~Pantic, ``Dimensional emotion prediction from spontaneous head
  gestures for interaction with sensitive artificial listeners,'' in {\em
  Intelligent virtual agents}, pp.~371--377, Springer, 2010.

\bibitem{jaiswal2016deep}
S.~Jaiswal and M.~F. Valstar, ``Deep learning the dynamic appearance and shape
  of facial action units,'' {\em Winter Conference on Applications of Computer
  Vision (WACV), 7-9 March 2016, Lake Placid, USA.}, 2016.

\bibitem{grimm2005evaluation}
M.~Grimm and K.~Kroschel, ``Evaluation of natural emotions using self
  assessment manikins,'' in {\em Automatic Speech Recognition and
  Understanding, 2005 IEEE Workshop on}, pp.~381--385, IEEE, 2005.

\bibitem{perez_gonzalez2010real}
E.~Perez-Gonzalez and J.~D. Reiss, ``A real-time semiautonomous audio panning
  system for music mixing,'' {\em EURASIP Journal on Advances in Signal
  Processing}, vol.~2010, no.~1, p.~436895, 2010.

\bibitem{nicolaou2011continuous}
M.~A. Nicolaou, H.~Gunes, and M.~Pantic, ``Continuous prediction of spontaneous
  affect from multiple cues and modalities in valence-arousal space,'' {\em
  IEEE Transactions on Affective Computing}, vol.~2, no.~2, pp.~92--105, 2011.

\bibitem{reisshighresaudio2016}
J.~D. Reiss, ``A meta-analysis of high resolution audio perceptual
  evaluation,'' {\em Journal of the Audio Engineering Society}, 2016.

\bibitem{yannakakis2011ranking}
G.~N. Yannakakis and J.~Hallam, ``Ranking vs. preference: a comparative study
  of self-reporting,'' in {\em International Conference on Affective Computing
  and Intelligent Interaction}, pp.~437--446, Springer, 2011.

\bibitem{schatz2012impact}
R.~Schatz, S.~Egger, and K.~Masuch, ``The impact of test duration on user
  fatigue and reliability of subjective quality ratings,'' {\em Journal of the
  Audio Engineering Society}, vol.~60, no.~1/2, pp.~63--73, 2012.

\bibitem{ackerman2009test}
P.~L. Ackerman and R.~Kanfer, ``Test length and cognitive fatigue: An empirical
  examination of effects on performance and test-taker reactions.,'' {\em
  Journal of Experimental Psychology: Applied}, vol.~15, no.~2, p.~163, 2009.

\bibitem{hafezi2015autonomous}
S.~Hafezi and J.~D. Reiss, ``Autonomous multitrack equalization based on
  masking reduction,'' {\em Journal of the Audio Engineering Society}, vol.~63,
  no.~5, pp.~312--323, 2015.

\bibitem{reiss2011intelligent}
J.~D. Reiss, ``Intelligent systems for mixing multichannel audio,'' in {\em
  Digital Signal Processing (DSP), 2011 17th International Conference on},
  pp.~1--6, IEEE, 2011.

\bibitem{ronan15subgrouping}
D.~Ronan, D.~Moffat, H.~Gunes, and J.~D. Reiss, ``Automatic subgrouping of
  multitrack audio,'' in {\em Proc. 18th International Conference on Digital
  Audio Effects {(DAFx-15)}}, 2015.

\bibitem{macompression}
Z.~Ma, B.~De~Man, P.~D. Pestana, D.~A.~A. Black, and J.~D. Reiss, ``Intelligent
  multitrack dynamic range compression,'' {\em Journal of the Audio Engineering
  Society}, 2015.

\end{thebibliography}
\bibliographystyle{ieeetr}

%
\vspace{-1.0cm}




\end{document}